\documentclass[useAMS,usenatbib,referee]{mn2e}
 
\usepackage{graphicx}
\usepackage{txfonts}

\def\lesssim{\mathrel{\hbox{\rlap{\hbox{\lower4pt\hbox{$\sim$}}}\hbox{$<$}}}}
\def\gtrsim{\mathrel{\hbox{\rlap{\hbox{\lower4pt\hbox{$\sim$}}}\hbox{$>$}}}}
\newcommand{\mincir}{\raise
-2.truept\hbox{\rlap{\hbox{$\sim$}}\raise5.truept
\hbox{$<$}\ }}
\newcommand{\magcir}{\raise
-2.truept\hbox{\rlap{\hbox{$\sim$}}\raise5.truept
\hbox{$>$}\ }}
\newcommand{\siml}{\raise -2.truept\hbox{\rlap{\hbox{$\sim$}}\raise5.truept
\hbox{$<$}\ }}
\newcommand{\simg}{\raise -2.truept\hbox{\rlap{\hbox{$\sim$}}\raise5.truept
\hbox{$>$}\ }}
\newcommand{\be}{\begin{equation}}
\newcommand{\ee}{\end{equation}}
\newcommand{\ba}{\begin{eqnarray}}
\newcommand{\ea}{\end{eqnarray}}

\newcommand {\ks} {km~s$^{-1} \;$}

\title{Shapley Optical Survey. I: Luminosity Functions in the
Supercluster Environment\footnote{Based on European Southern
Observatory Archive Data.}}  

\author[A. Mercurio et al.]
{A. Mercurio$^1$, P. Merluzzi $^1$, C. P. Haines $^1$, A. Gargiulo
$^1$, N. Krusanova$^1$, 
\newauthor
G. Busarello$^1$, F. La Barbera $^1$, M. Capaccioli$^{1,2,3}$, G. Covone$^1$ \\
$^{1}$ INAF - Osservatorio Astronomico di Capodimonte, via Moiariello
16, I-80131 Napoli, Italy \\
$^{2}$ Physics Department, Universit\`a degli Studi ``Federico II'',
Napoli, Italy \\
$^{3}$ VST Center at Naples (VSTceN), via Moiariello
16, I-80131 Napoli, Italy }

\begin{document}

\date{Accepted. Received}

\pagerange{\pageref{firstpage}--\pageref{lastpage}} \pubyear{2005}

\maketitle

\label{firstpage}

\begin{abstract}

We present the Shapley Optical Survey, a photometric study covering a
$\sim$ 2 deg$^2$ region of the Shapley Supercluster core at
z$\sim$0.05 in two bands (B and R). The galaxy sample is complete to B
= 22.5 ($>$M$^*$+6, $\mathrm{N_{gal}}$ = 16\,588), and R = 22.0
($>$M$^*$+7, $\mathrm{N_{gal}}$ = 28\,008). The galaxy luminosity
function cannot be described by a single Schechter function due to
dips apparent at B $\sim$ 17.5 (M$_\mathrm{B} \sim$ - 19.3) and R
$\sim$ 17.0 (M$_\mathrm{R} \sim$ - 19.8) and the clear upturn in the
counts for galaxies fainter than B and R $\sim 18$ mag.  We find,
instead, that the sum of a Gaussian and a Schechter function, for
bright and faint galaxies respectively, is a suitable representation
of the data.  We study the effects of the environment on the
photometric properties of galaxies, deriving the galaxy
luminosity functions in three regions selected according to the local
galaxy density, and find a marked luminosity segregation, in the
sense that the LF faint-end is different at more than 3$\sigma$
confidence level in regions with different densities.  In addition,
the luminosity functions of red and blue galaxy populations show very
different behaviours: while red sequence counts are very similar to
those obtained for the global galaxy population, the blue galaxy
luminosity functions are well described by a single Schechter function
and do not vary with the density.  Such large
environmentally-dependent deviations from a single Schechter function
are difficult to produce solely within galaxy merging or suffocation
scenarios. Instead the data support the idea that mechanisms related
to the cluster environment, such as galaxy harassment or ram-pressure
stripping, shape the galaxy LFs by terminating star-formation and
producing mass loss in galaxies at $\sim{\mathrm M}^*+2$, a magnitude
range where blue late-type spirals used to dominate cluster
populations, but are now absent.
\end{abstract}


\section{Introduction}
\label{intro}

The properties and evolution of galaxies are strongly dependent on
environment (e.g., Blanton et al. \citeyear{bla05}; Rines et
al. \citeyear{rin05}; Smith et al.  \citeyear{smi05}; Tanaka et
al. \citeyear{tan05}). In particular the cluster galaxy population has
evolved rapidly over the last 4 Gyr (Butcher \& Oemler
\citeyear{but84}). While distant clusters are dominated, particularly
at faint magnitudes, by blue spiral galaxies, often with signs of
disturbed morphologies and evidence of multiple recent star-formation
events (Dressler et al. \citeyear{dre94}), local clusters are
completely dominated by passive early-type galaxies.

Recent observational studies on the luminosity, colours, morphology
and spectral properties of galaxies have pointed out that the physical
mechanisms which produce the transformation in galaxies affecting both
the structure and the star formation are naturally driven by and
related to the environment (e.g., Treu et al. \citeyear{tre03}). These
processes are linked in various ways to the local density and the
properties of the intra-cluster medium (ICM).  In fact, galaxy-ICM
interactions, such as ram-pressure stripping and suffocation, require
a dense ICM and take place principally in the central cluster
regions. The high density regions are also characterized by a steep
cluster potential, and we can expect that galaxy-cluster gravitational
interactions such as tidal stripping and tidal triggering are also
dominant. On the contrary, in low-density environment galaxies have
never been through the cluster centre and therefore have never
experienced the effects of tidal stripping and tidal triggering of
star formation, so the dominant mechanisms are galaxy-galaxy
interactions, in terms of both low-speed interactions between galaxies
of similar mass (mergers) and high-speed interactions between galaxies
in the potential of the cluster (harassment).

The above mentioned mechanisms affect differently the observed
morphology and/or the star formation properties of galaxies. In
particular, galaxy harassment and ram pressure stripping cause a
partial loss of gas mass and, depending on the fraction of gas removed
and its rate, ram-pressure stripping can lead either to a rapid
quenching of star formation or to a slow decrease in the star
formation rate (Larson, Tinsley, \& Caldwell \citeyear{lar80}; Balogh
et al. \citeyear{bal00}; Diaferio et al. \citeyear{dia01}; Drake et
al. \citeyear{dra00}).  Successive high-speed encounters between
galaxies (galaxy harassment) lead to gas inflow and strong star
formation activity (Fujita \citeyear{fuj98}).

With the aim to investigate the effects of the environment on the
galaxy population, we have undertaken an optical study of the Shapley
Supercluster (SSC) core, one of the densest structures in the nearby
Universe. The study of this region, selected because of its physical
peculiarity in terms of density and complexity, but also for the
availability of multiwavelength observations, will take advantage of
deep optical photometry from the ESO Archive covering an area of 2
deg$^2$.  The Shapley Optical Survey (SOS) in B and R bands provides a
galaxy sample complete and reliable up to 22.5 mag and 22.0 mag in B
and R bands, respectively.  We plan to use the excellent SOS dataset
to study, with respect to the supercluster environment, the
distribution of galaxy populations both in luminosity and colour and
the galaxy structural properties comparing the observations with
theoretical predictions. In the present paper we will present the
dataset, the catalogues and the galaxy luminosity functions in B and R
bands as function of the SSC environment.

The galaxy luminosity function (LF), which describes the number of
galaxies per unit volume as function of luminosity, is a powerful tool
to constrain galaxy transformations, since it is directly related to
the galaxy mass function. Moreover, the effect of environment on the
observed galaxy LF could provide a powerful discriminator among the
proposed mechanisms for the transformations of galaxies. The effects
of galaxy merging and suffocation on the cluster galaxy population
have been studied through combining high-resolution N-body simulations
with semi-analytic models for galaxy evolution (e.g., Springel et
al. \citeyear{spi01}; Kang et al. \citeyear{kan05}). These show that
while galaxy merging is important for producing the most luminous
cluster galaxies, the resultant LF can always be well described by a
Schechter (Schechter \citeyear{sch76}) function, although both M$^*$
and the faint-end slope can show mild trends with environment. Galaxy
mergers are also inhibited once the relative encounter velocities
become much greater than the internal velocity dispersion of galaxies,
and so are rare in rich clusters (Ghigna et al. \citeyear{ghi98}).
In contrast, galaxy harassment and ram-pressure stripping may
change the LF shape as galaxies lose mass in interactions with other
galaxies, the cluster's tidal field, and the ICM. In particular
\citeauthor{moo98} (\citeyear{moo98}) showed that harassment has
virtually no effect on a system as dense as a giant elliptical galaxy
or a spiral bulge and only purely disk galaxies can be turned into
spheroidals, so these mechanisms produces a cutoff for Sd-Im
galaxies. Since the luminosity function is strongly type
specific, and those for Sc and Sd/Im galaxies can be described by
narrow ($\sigma\sim1$ mag) Gaussian distributions centred at $\sim$
M$^*+$1 and $\sim$ M$^*+3$ (de Lapparent \citeyear{del03bis}), 
the effects of galaxy harassment could be characterized by a dip in
the LF at these magnitudes.

In order to further investigate and to assess the relative importance
of the processes that may be responsible for the galaxy
transformations, we have performed a photometric study of the SSC
core, examining in particular the effect of the environment through
the comparison of luminosity functions in regions with different local
densities.

The SSC was observed by Raychaudhury (\citeyear{ray89}) and
the LF was firstly derived by Metcalfe, Godwin \& Peach
(\citeyear{met94}; hereafter MGP94). By using photographic data, they
investigate a region of 4.69 deg$^2$ around the cluster A\,3558
considering a sample of 4599 galaxies complete and uncontamiminated by
stars (to 2$\%$ level) for ${\it b}<19.5$. The derived LF for the
central region of 1.35 deg$^2$ showed a broad peak in the number of
galaxies at ${\it b}=18$ which cannot be well fitted by a Schechter
function. Moreover, MGP94 found a deficit of blue galaxies in the
A\,3558 core suggesting morphological segregation. However, their
study is limited to bright magnitudes, preventing the determination of
the faint-end slope while taking advantage of deeper photometry and
larger sample of galaxies distributed in larger SSC area, we can
provide clear evidence on the LF shape thus quantifying the
environmental effect on the LF properties.

The layout of this work is the following. General information of the
structure of the SSC core are summarized in Sect.~\ref{sec:SSC}. We
describe observations, data reduction and the photometric calibrations
in Sect.~\ref{sec:2}. The catalogues are presented in
Sect.~\ref{sec:4}. Sect.~\ref{sec:5} is dedicated to the definition of
the environment and in and Sect.~\ref{sec:6} we show the LFs. Finally
Sect.~\ref{sec:7} contains the summary and the discussion of the
results. In this work we assume H$_0$ = 70 \ks\,Mpc$^{-1}$, $\Omega_m$
= 0.3, $\Omega_{\Lambda}$ = 0.7. According to this cosmology, 1 arcmin
corresponds to 0.060 Mpc at $z=0.048$.

\section{The Shapley Supercluster}
\label{sec:SSC}

\begin{figure*}
\begin{center}
\centerline{\resizebox{18cm}{!}{\includegraphics{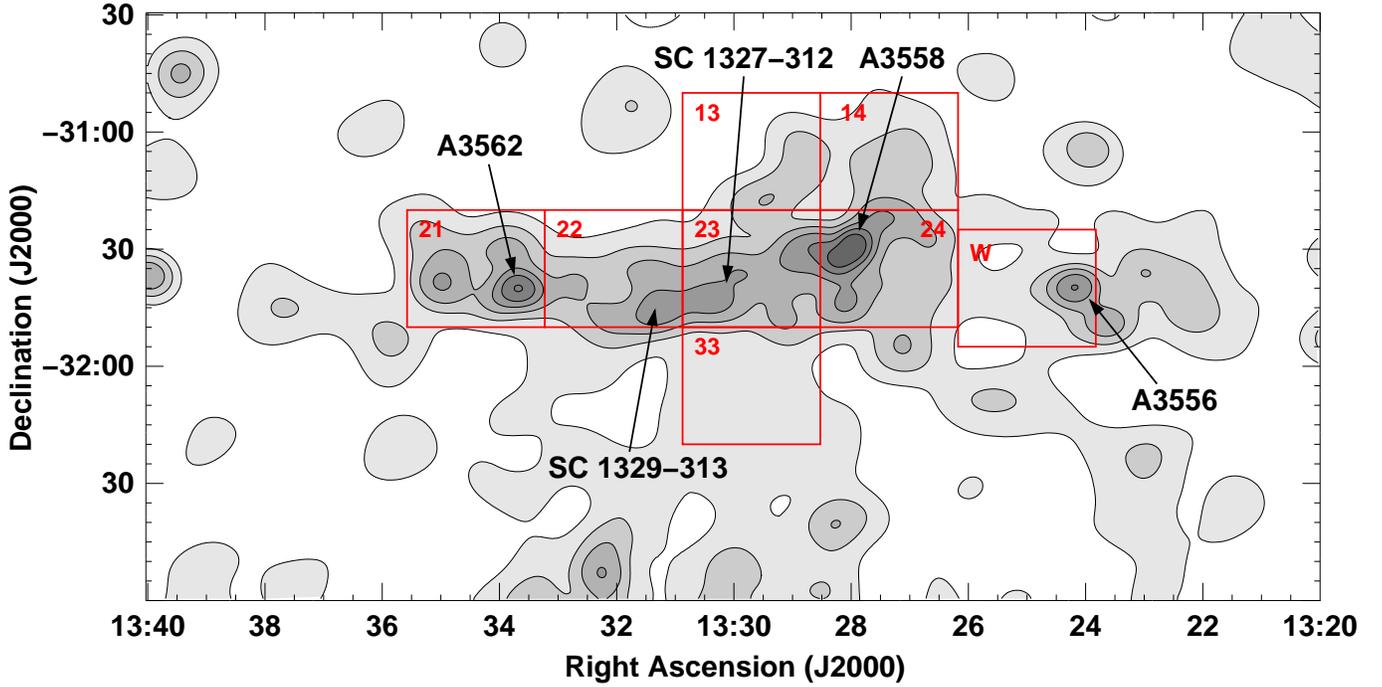}}}
\caption{The surface density of R$\mathrm{<}18.5$ galaxies of the the SSC
core, obtained by using data of the SuperCosmos Sky
Survey (\citeauthor{ham01} \citeyear{ham01}). Red rectangles indicate the 8
analysed fields in SOS.}
\end{center}
\label{fig1}
\end{figure*}

The SSC represents an ideal target for the investigation of the role
played by environment in the transformation of galaxies, and has been
investigated by numerous authors since its discovery (Shapley
\citeyear{sha30}).  It is one of the richest supercluster in the
nearby universe, consisting of as many as 25 Abell clusters in the
redshift range $0.035<$z$<0.055$. Extensive redshift surveys (Bardelli
et al. \citeyear{bar00}; Quintana, Carrasco \& Reisenegger
\citeyear{qui00}; Drinkwater et al. \citeyear{dri04}) indicate that
these clusters are embedded in two sheets extending over a
$\sim10\times20$\,deg$^2$ region of sky ($\sim35\times$70
$h_{70}^{-2}$\,Mpc$^2$), and that as many as half the total galaxies
in the supercluster are from the inter-cluster regions. The Shapley
core (Fig.~\ref{fig1}) is constituted by three Abell clusters: A\,3558
(z=0.048, Melnick \& Quintana \citeyear{mel81}; Metcalfe, Godwin \&
Spenser \citeyear{met87}; Abell richness R=4, Abell, Corwin \& Olowin
\citeyear{abe89}), A\,3562 (z=0.049, \citeauthor{str99}
\citeyear{str99}, R=2, Abell et al. \citeyear{abe89}) and A\,3556
(z=0.0479, \citeauthor{str99} \citeyear{str99}, R=0, Abell et
al. \citeyear{abe89}) and two poor clusters SC\,1327-312 and SC
1329-313. Dynamical analysis indicates that at least a region of
radius 11$\,h_{70}^{-1}$\,Mpc centred on the central cluster A\,3558,
and possibly the entire supercluster, is past turnaround and is
collapsing (Reisenegger et al. \citeyear{rei00}), while the core
complex itself is in the final stages of collapse, with infall
velocities reaching $\sim$2000\,km\,s$^{-1}$.

A major study of the dynamical properties of the supercluster core was
performed by Bardelli et al. (\citeyear{bar01} and reference
therein). They showed that the supercluster core has a complex, highly
elongated structure, and identified 21 significant 3-dimensional
subclumps, including eight in the A\,3558 cluster alone.

The X-ray observations show that the supercluster has a flattened and
elongated morphology where clusters outside the dense core are
preferentially located in hot gas filaments (Bardelli, Zucca, \&
Malizia \citeyear{bar96}; Kull \& B$\mathrm{\ddot{o}}$hringer
\citeyear{kul99}; \citeauthor{def05} \citeyear{def05}). Moreover,
Finoguenov et al. (\citeyear{fin04}) showed a strong interaction
between the cluster A\,3562 and the nearby group SC\,1329-313 with an
associated radio emission having young age (Venturi et
al. \citeyear{ven00}, \citeyear{ven03}). However, since this is one of
the weakest radio halos found, Venturi et al. (\citeyear{ven03})
suggested that this halo is connected with the head-on radio galaxy of
A\,3562.  Bardelli et al. (\citeyear{bar01}) suggested that the
A\,3558 complex is undergoing a strong dynamical evolution through
major merging seen just after the first core-core encounter, and so
the merging event has already been able to induce modifications in the
galaxy properties. Very recently, Miller (\citeyear{mil05}), with a
radio survey of a 7 deg$^2$ region of SSC, found a dramatic increase
in the probability for galaxies in the vicinity of A\,3562 and
SC\,1329-313 to be associated with radio emission. He interpreted this
fact as a young starburst related to the recent merger of SC\,1329-31
with A\,3562.

\section{Observations, Data reduction and photometric calibration}
\label{sec:2}

The SOS data obtained from the ESO Archive (68.A-0084, P.I. Slezak),
were acquired with the ESO/MPI 2.2-m telescope at La Silla. We
analysed B- and R-band photometry of eight contiguous fields covering
a 2 deg$^2$ region centred on the SSC, as shown in Fig. 1.

The observations (see Table~\ref{obs} for details) were carried out
with the WFI camera, a mosaic of eight 2046 $\times$ 4098 pixels CCDs,
mounted on the Cassegrain focus of the telescope. The camera has a
field of view of 34$^\prime$ $\times$ 33$^\prime$, corresponding to
2.0 $\times$ 1.9 h$_{70}^{-2}$ Mpc$^{2}$ at the cluster redshift,
and a pixel scale of 0.238 arcsec.  The total exposure times for each
field are 1500 s (300 s $\times$ 5) in B band and 1200 s (240 s
$\times$ 5) in R band, reaching the R=25 (B=25.5) at 5$\sigma$. The
single exposures are jittered to cover the gaps between the different
CCDs of the camera. Landolt (\citeyear{lan92}) stars were observed in
order to perform accurate photometric calibration.

\begin{table}
\begin{center}
\caption{The observations.}
  \label{obs}
\begin{tabular}{c c c c c}
  \hline
  \hline
  Field & Band & Centre & Date & FWHM \\
   \#     &      & RA, Dec&      & arcsec \\
  \hline
  \hline
   21 & B & 13:34:24.1, -31:34:57.1 & 18 March 2002 & 0.95\\
    & R & 13:34:24.1, -31:34:57.0 &      ''       & 0.87\\
   22 & B & 13:32:03.2, -31:34:57.1 & 18 March 2002 & 0.79\\
    & R & 13:32:03.2, -31:34:57.1 &      ''       & 0.70\\
   23 & B & 13:29:42.4, -31:34:57.3 & 18 March 2002 & 0.76\\
    & R & 13:29:42.4, -31:34:57.5 &      ''       & 0.71\\
   24 & B & 13:27:21.5, -31:34:57.4 & 18 March 2002 & 0.77\\
    & R & 13:27:21.5, -31:34:57.1 &      ''       & 0.73\\
   W  & B & 13:25:00.6, -31:40:27.1 & 18 March 2002 & 0.83\\
    & R & 13:25:00.6, -31:40:26.6 &      ''       & 0.73\\
   13 & B & 13:29:42.5, -31:04:58.0 &  9 April 2003 & 0.98\\
    & R & 13:29:42.5, -31:04:59.3 &  9 June  2002 & 1.11\\
   14 & B & 13:27:21.6, -31:04:57.6 & 19 March 2002 & 1.16\\
    & R & 13:27:21.5, -31:04:56.6 &      ''       & 1.27\\
   33 & B & 13:29:42.4, -32:04:57.5 &  9 April 2003 & 0.81\\
      & R & 13:29:42.3, -32:04:58.9 &  9 June  2002 & 1.43\\
  \hline
\end{tabular}
\end{center}
\end{table}

We used the ALAMBIC pipeline (version 1.0, Vandame \citeyear{van04}) to
reduce and combine the SOS images. The pipeline follows the standard
procedures for bias subtraction and flat-field correction; the
twilight sky exposures for each band were used to create the master
flat.

The photometric calibration was performed into the
Johnson-Kron-Co\-u\-sins photometric system using the Landolt
stars. With the IRAF tasks DAOPHOT and APPHOT we computed the
instrumental magnitude of the stars in a fixed aperture in B and R
bands.

For the R band we calibrated the flux by adopting the following
relation:

\begin{equation}
    M' = M + \gamma C + AX + ZP,
\end{equation}
where $M$ is the magnitude of the star in the standard system, $M'$ is
the instrumental magnitude, $\gamma$ is the coefficient of the colour
term, $C$ is the colour of the star in the standard system, $A$ is the
extinction coefficient, $X$ is the airmass and $ZP$ is the zero
point. For the B band we took into account the colour term
(B-R). The results are reported in Table~\ref{tab2}.

\begin{table*}
\begin{center}
\caption{The results of the photometric fit for B and R band.}
\begin{tabular}{c c c c c c c}
  \hline
  \hline
  Observing Night & Band & $C$ &$ZP$ & $A$ & $\gamma$ & rms \\
  \hline
  \hline
  18 March 2002 & B & B - R & -24.531 $\pm$ 0.024 & 0.189 $\pm$ 0.017& -0.131 $
\pm$ 0.009 & 0.041 \\
  18 March 2002 & R & V - R & -24.548 $\pm$ 0.038 & 0.147 $\pm$ 0.026&  0.049 $
\pm$ 0.035 & 0.036 \\
  9 April 2003  & B & B - R & -24.562 $\pm$ 0.029 & 0.162 $\pm$ 0.019& -0.141 $
\pm$ 0.017 & 0.034 \\
  \hline
\label{tab2}
\end{tabular}
\end{center}
\end{table*}
Since photometric standards were not available for the nights of 9 June
2002 and 19 March 2002, we adopted relative calibration for fields \#13,
\#33 in R band and field \#14 both in B and R band. The photometric
accuracy for the zero point was about 0.04 mag in both bands.

\section{The catalogues}
\label{sec:4}

The photometric catalogues from the SOS images were produced using
SExtractor (Bertin \& Arnouts, \citeyear{ber96}) together with a set
of software procedures developed by the authors in order to increase
the quality of final catalogues, avoiding spurious detections and
misleading results (see Sect.~\ref{sec:41}).

The star/galaxy classification was based on both the parameter {\it
class star} (CS) of SExtractor and the value of the {\it full width at
half maximum} (FWHM). Stars were defined as those objects with
CS$\ge$ 0.98 or having FWHM equal to  those of bright, non satured
stellar sources in the image.

The completeness magnitudes were firstly estimated using the
prescription of Garilli et al. (\citeyear{gar99}). Then the
reliability and the completeness of the catalogues were checked
performing Montecarlo simulations by adding artificial stars and
galaxies (see Sect.~\ref{sec:42}). The final catalogues consist of
16\,588 and 28\,008 galaxies in B and R band, respectively, up to the
completeness magnitude limits B=22.5 and R=22.0.

Aperture and Kron (Kron \citeyear{kro80}) magnitudes were measured in
each band. For aperture photometry we referred to the aperture of 17
arcsec ($\sim$ 8 kpc) of diameter used by Bower, Lucey and Ellis
(\citeyear{bow92}) for Coma.  Converting this value from Coma redshift
to our redshift we used an aperture of 8 arcsec of diameter.  Kron
magnitudes ($\mathrm{M}_{Kron}$) were computed in an adaptive aperture
with diameter $a \cdot \mathrm{R}_{Kron}$, where $\mathrm{R}_{Kron}$
is the Kron radius and $a$ is a constant. We chose $a$ = 2.5, yielding
$\sim$ 94\% of the total source flux within the adaptive aperture
(Bertin \& Arnouts \citeyear{ber96}).  We measured the Kron magnitude
for all the objects in the catalogues without applying any correction
to the total magnitude.  The uncertainties on the magnitudes were
obtained by adding in quadrature both the uncertainties estimated by
SExtractor and the uncertainties on the photometric calibrations.  The
measured magnitudes were corrected for galactic extinction (B=0.238
and R=0.149) following \citeauthor{sch98}
(\citeyear{sch98}). Luminosity functions were computed by means of
Kron magnitudes, while aperture magnitudes were used for measuring
galaxy colours. SOS catalogues are available on request.

\subsection{Cleaning procedure}
\label{sec:41}

In order to obtain {\it clean} catalogues we used the following approach
that takes into account both the performances of SExtractor and the
characteristics of the analysed fields (crowdness, background
fluctuations, bright objects sizes and distribution).

We ran SExtractor with two different deblending parameters. We produced
the bulk of the catalogue adopting a low deblending parameter
(0.0001), which allows a suitable detection of close objects. Then we
corrected the multiple detections of bright extended objects using a
high deblending parameter ($>$0.01).

The combined images show a significant number of bad and warm pixels,
and cosmic rays residuals often detected by SExtractor as
sources. These spurious detections were identified and then removed
since either they are present only in few exposures or they are
particularly compact, comparing their $\mathrm{M}_{Kron}$ with the
magnitude measured over the central pixel.

In the vicinity of bright galaxies with extended halos, SExtractor
sometimes overestimates R$_{Kron}$ and M$_{Kron}$.  These objects were
identified and their M$_{Kron}$ corrected.

Finally, we removed spurious objects, ghosts or diffraction spikes
around bright (R$<$15) stars by defining circular avoidance regions
(whose area is proportional to the stars flux level).

\subsection{Completeness and reliability}
\label{sec:42}

The first estimates of the completeness magnitudes, derived using
the prescription of Garilli et al. (\citeyear{gar99}), are 23.0 in the
R band, and 23.5 in the B band.  We checked the reliability and
completeness of the SOS catalogues for each 0.5 magnitude bin by
adding 10\,000 artificial stars and galaxies to the images, and
computing the fraction of these sources detected and correctly
classified by SExtractor. The artificial stars were created by taking
a bright, non-saturated star ($\mathrm{R}\sim17$) in the image and
dimming it to the appropriate magnitude, while the galaxies were
simulated by taking galaxies of differing Hubble types and the
appropriate magnitude from the Hubble Ultra Deep Field (using
photometry from the COMBO-17 Chandra Deep Field South catalogue; Wolf
et al. \citeyear{wol04}), resampling them to the resolution of the
WFI, and convolving them with the image Point Spread Function (PSF).

\begin{table}
\begin{center}
\caption{Completeness and reliability of the SOS catalogues}
\begin{tabular}{ccc}
  \hline
R & completeness & \% of stars \\
  \hline
  \noalign{\smallskip}
mag & & missclassified \\ \hline
20.0--20.5 & 98.4 & 2.1 \\
20.5--21.0 & 97.3 & 3.0 \\
21.0--21.5 & 95.3 & 4.1 \\
21.5--22.0 & 94.3 & 8.0 \\
22.0--22.5 & 92.0 & 34.8 \\
22.5--23.0 & 88.1 & 75.6 \\ \hline
\end{tabular}
\label{completeness}
\end{center}
\end{table}

At $\mathrm{R}=22.0$, 94.3\% of the simulated galaxies were
successfully detected and classified. In the SOS field the number
of stars and galaxies become equal at $\mathrm{R}=21.4$. Beyond
$\mathrm{R}=22.0$ the fraction of stars misclassified as galaxies
increases dramatically mainly due to the blending of the
sources.  Moreover, further stellar contamination is due to the high
number density of both stars and galaxies in this field (the Galactic
latitude of the field is $+30^{\circ}$) which increases the frequency
of star-star and star-galaxy blends that can be misclassified as
single galaxies. The estimates of completeness and reliability for the
R band are shown in Table~\ref{completeness}. Analogous results were
obtained for the B band.

We adopted the conservative limits $\mathrm{R}=22.0$ and $\mathrm{B}=22.5$ as
the magnitudes below which stellar contamination can be modelled and
accounted for in the galaxy LF determination.

\section{Quantifying the Galaxy environment}
\label{sec:5}

\begin{figure*}
\centerline{{\resizebox{18cm}{!}{\includegraphics{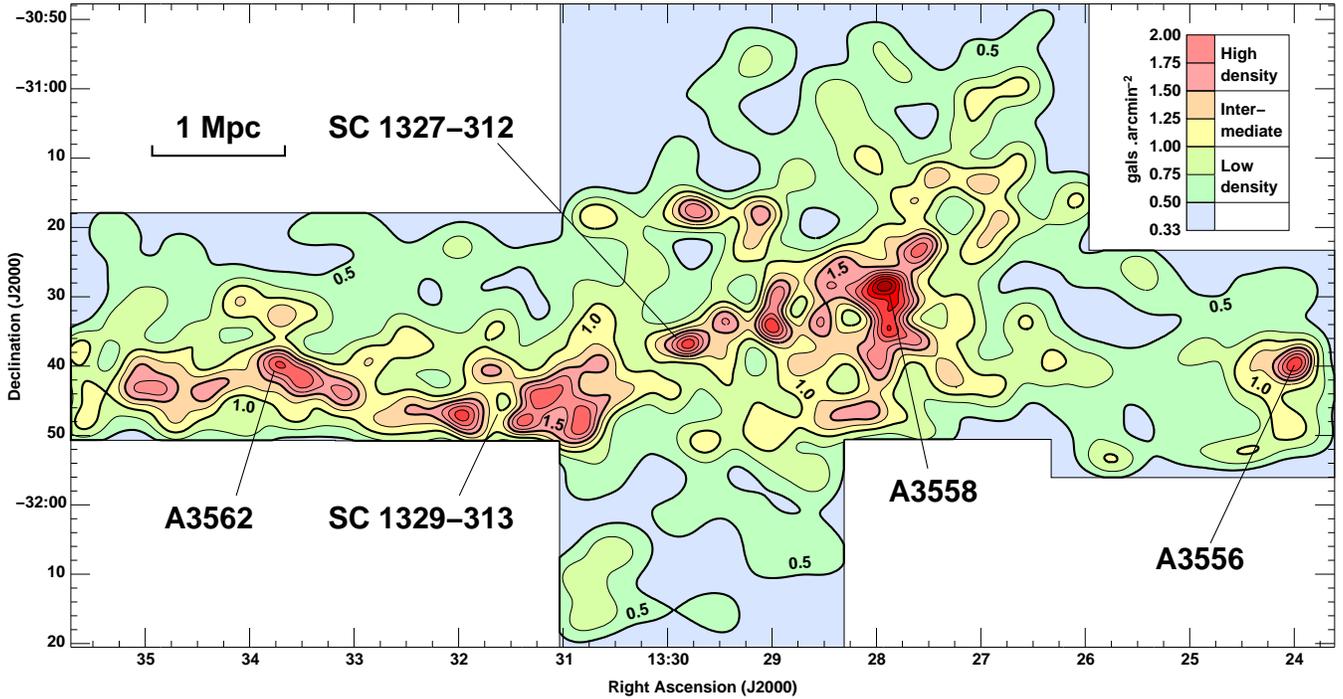}}}}
\caption{The surface density of $\mathrm{R}<21.0$ galaxies in the region of the
  SSC core complex. Isodensity contours are shown at
  intervals of 0.25 galaxies arcmin$^{-2}$, with the thick contours 
corresponding to 0.5, 1.0 and 1.5 galaxies arcmin$^{-2}$, the
  densities used to separate the three cluster environments. 
The area corresponds to 9.0$\times$5.4 h$_{70}^{-2}$ Mpc$^2$ at z=0.048.} 
\label{densitymap}
\end{figure*}

To study the effect of the environment on galaxies in the SSC core,
the local density of galaxies, $\Sigma$, was determined across the WFI
mosaic. This was achieved using an adaptive kernel estimator (Pisani
\citeyear{pis93}; \citeyear{pis96}), in which each galaxy $i$ is
represented by a Gaussian kernel,
$K(r_i)\propto\exp(-r^{2}/2\sigma_{i}^{2})$, whose width $\sigma_{i}$
is proportional to $\Sigma_{i}^{-1/2}$ thus matching the resolution
locally to the density of information available. For this study, we
considered the surface number density of R$<21.0$
($<\mathrm{M^{*}}+6$) galaxies, with an additional colour cut applied
to reject those galaxies more than 0.2 mag redder in B-R than the
observed red sequence (Eq.~\ref{CM}) to minimize background
contamination. As there are no known structures in the foreground of
the SSC core (90\% of R$<16$ galaxies have redshifts confirming that
they belong to the supercluster), any substructure identified in the
density map is likely to be real and belonging to the supercluster.
The local density was initially determined using a fixed Gaussian
kernel of width 2 arcmin, and then iteratively recalculated using
adaptive kernels. The resultant surface density map of the SSC core
complex is shown in Fig.~\ref{densitymap}, with the three clusters and
two groups indicated. Isodensity contours are shown at intervals of
0.25 galaxies arcmin$^{-2}$, with the thick contours corresponding to
0.5, 1.0 and 1.5 galaxies arcmin$^{-2}$, the densities used to
separate the three supercluster environments described below. The
background density of galaxies as estimated from an area of
5\,deg$^{2}$ of the Deep Lens Survey (DLS, Wittman et
al. \citeyear{wit02}) is 0.335 galaxies arcmin$^{-2}$, and hence the
thick contours correspond to overdensity levels of $\sim50$, 200 and
400 galaxies $h_{70}^{2}\,$Mpc$^{-2}$ respectively. The entire region
covered by the SOS is overdense with respect to field galaxy
counts.

\section{SOS Luminosity Functions}
\label{sec:6}

The LF in each band in the whole surveyed area was obtained up to the
completeness magnitude limits, removing the interlopers by
statistically subtracting the background contamination, as determined
from thirteen control fields of the DLS, covering a total area of
$\sim$ 4.4 deg$^2$. The adopted procedures for background subtraction
are explained in Sect.~\ref{sec:61}.

In order to investigate the effects of the environments we
derived and compared the LFs, determined in the three different
regions characterized by high-, intermediate- and low-densities
of galaxies.

We fitted the observed galaxy counts with a single Schechter (S)
function. However, since the LFs were generally poorly fitted by using
such a model, the fits were also computed with the sum of Gaussian and
Schechter (G+S) functions (Sect.~\ref{sec:62} and Sect.~\ref{sec:63})
in order to describe bright and faint galaxy populations (e.g., de
Lapparent et al. \citeyear{del03}; Molinari et al. \citeyear{mol98};
Biviano et al. \citeyear{biv95}).  Moreover, we compared the LFs with
the counts of the red sequence galaxies (Sect.~\ref{sec:64}). We
selected also galaxies bluer than the colour magnitude relation in
order to derive the luminosity function of late-type galaxy
populations.

Absolute magnitudes were determined using the k-corrections for
early-type galaxies from models of Bruzual \& Charlot
(\citeyear{bru03}). All the fit parameters and associated $\chi^{2}$
statistics are listed in Table~\ref{fitsLF}.

   \begin{table*} 
     \caption[]{Fits to the LFs. Errors on the
      $\mathrm{M^*}$ and $\alpha$ parameters can be obtained from the
      confidence contours shown in Figs.~\ref{totLF}, \ref{contB}
      and \ref{contR}. In the table S indicates the fit with a single
      Schechter and G+S those with Gaussian plus Schechter.}

     $$
           \begin{array}{c c c | c c c | c c c | c c }
            \hline
            \noalign{\smallskip}
	    \mathrm{Region} &   \mathrm{Band} & \mathrm{Function} &
            \mathrm{m^*} & \mathrm{M^*} &\alpha & \multicolumn{2}{c}{\mu}  & \sigma  & \mathrm{\chi^2_{\nu}} & \mathrm{P(\chi^2>\chi^2_{\nu})}\\
		\noalign{\smallskip}
		\hline
		\hline
	    \noalign{\smallskip}
            \mathrm{all \ \ field} & \mathrm{B} & \mathrm{S} & 15.35 & -21.42 & -1.46 & & & & 2.62 & 0.08\% \\
            \mathrm{all \ \ field}  & \mathrm{B} & \mathrm{G+S} & 15.53 & -21.24 & -1.74 & 17.01 & -19.76 & 1.32 & 0.36& 97.1\% \\
            \mathrm{high \ \ density} &  \mathrm{B} & \mathrm{S} & 14.64 & -22.13 & -1.46 & & & & 0.95& 50.3\% \\
            \mathrm{high \ \ density} & \mathrm{B} & \mathrm{G+S} & 16.47 & -20.32 & -1.51 & 17.00 & -19.77 & 1.73 & 0.94&50.0\% \\
            \mathrm{int \ \ density}  & \mathrm{B} & \mathrm{S} & 15.01 & -21.76 & -1.50 & & & & 1.33& 18.6\% \\
            \mathrm{int \ \ density} & \mathrm{B} & \mathrm{G+S} & 15.46 & -21.31 & -1.56 & 16.51 & -20.26 & 0.85 & 0.50& 89.1\% \\
            \mathrm{low \ \ density} & \mathrm{B} & \mathrm{S} & 15.57 & -21.20 & -1.49 & & & & 2.22 & 0.69\% \\
            \mathrm{low \ \ density} & \mathrm{B} & \mathrm{G+S} & 16.11 & -20.66 & -1.66 & 16.96 & -19.81 & 1.09 & 0.60& 81.5\% \\
            \mathrm{all \ \ field} & \mathrm{R} & \mathrm{S} & 14.52 & -22.26 & -1.26 & & & & 1.23& 23.5\% \\
            \mathrm{all \ \ field} & \mathrm{R} & \mathrm{G+S} & 13.72 & -23.06 & -1.62 & 15.89 & -20.89 & 1.23 & 0.46& 94.7\% \\
            \mathrm{high \ \ density} & \mathrm{R} & \mathrm{S} & 14.29 & -22.49 & -1.30 & & & & 0.86& 61.7\% \\
            \mathrm{high \ \ density} & \mathrm{R} & \mathrm{G+S} & 14.15 & -22.63 & -1.30 & 20.92 & -15.86 & 3.15 & 1.01& 43.8\% \\
            \mathrm{int \ \ density} & \mathrm{R} & \mathrm{S} & 14.27 & -22.51 & -1.39 & & & & 1.28& 20.5\% \\
            \mathrm{int \ \ density} & \mathrm{R} & \mathrm{G+S} & 15.00 & -21.78 & -1.43 & 15.22 & -21.49 & 0.98 & 0.74& 71.3\% \\
            \mathrm{low \ \ density} & \mathrm{R} & \mathrm{S} & 13.75 & -23.03 & -1.50 & & & & 3.33& 0.002\% \\
            \mathrm{low \ \ density} & \mathrm{R} & \mathrm{G+S} & 15.28 & -21.50 & -1.80 & 16.37 & -20.41 & 1.58 & 1.25& 24.7\% \\
            \noalign{\smallskip}
            \hline
            \noalign{\smallskip}
            \hline
         \end{array}
    $$
      \label{fitsLF}

   \end{table*}

   \begin{figure} 
   \centerline{\resizebox{\hsize}{!}{\includegraphics[angle=-90]{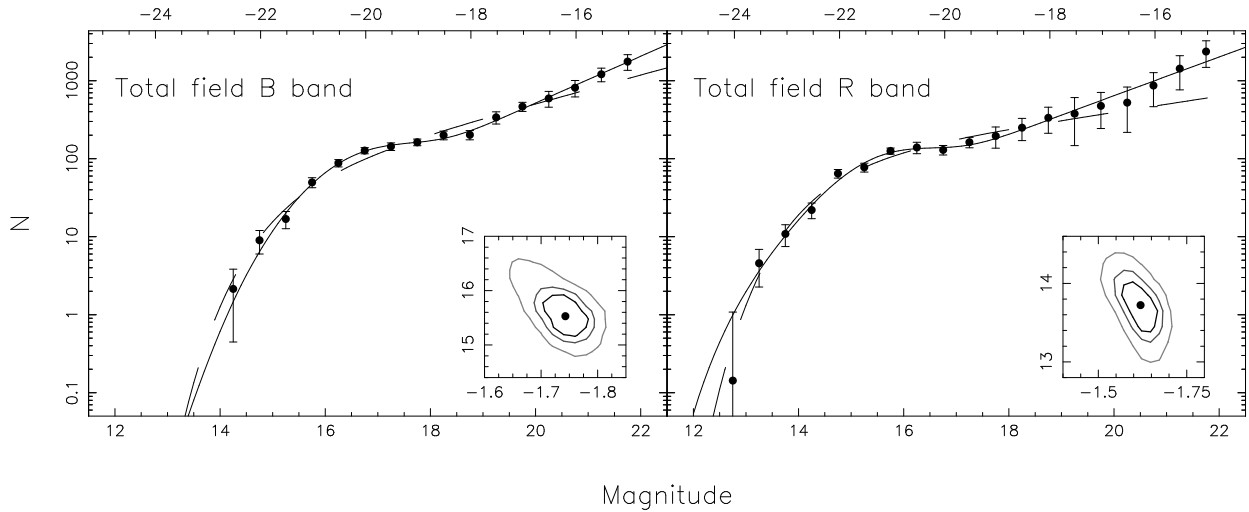}}}
   \caption{Luminosity function in the B and R bands over 2 deg$^2$
   field covering the core of the SSC.  Dashed and continuous lines
   are fits with the S and with the G+S functions (see text),
   respectively.  In the small panels the 1, 2 and 3$\sigma$
   confidence levels of the best--fit parameters for $\alpha$ and
   M$^*$ from the G+S fit, are shown. 
   The counts are per half magnitudes.}
   \label{totLF} 
   \end{figure}

\subsection{Background galaxy subtraction}
\label{sec:61}

Since the region covered by the SOS lies completely within the
overdensity corresponding to the core complex, to perform a reliable
statistical subtraction of field galaxies a suitable large control
field is required, that has been observed with a similar filter set (B
and R) to at least the depth of the SOS. The large area is
necessary in order to minimize the effects of cosmic variance and
small number statistics.

To this aim we chose the DLS, which consists of deep BVRz' imaging of seven
$2^{\circ}\times2^{\circ}$ degree fields. The observations have been
made using the Mosaic-II cameras on the NOAO KPNO and CTIO 4-m
telescopes, with exposure times of 12\,000s in BVz' and 18\,000 s in
R, resulting in 5$\sigma$ depths of B,V,R$\sim27$. The R-band
images were obtained in good seeing conditions with a FWHM of 0.9$''$
whereas the other bands have FWHM around 1.2$''$.

The catalogues were extracted following the same procedures of the SOS
data. We considered thirteen $35^{\prime}\times35^{\prime}$ Mosaic-II
fields, covering a total area of $\sim$ 4.44 square degrees (after removal
of regions around bright stars) in two well separated regions of sky
(fields 2 and 4 in the DLS). Given the depth of the DLS images,
star-galaxy separation using the combined stellarity-FWHM
classification method was found to be $>$99\% efficient to
$\mathrm{R}=22.0$. There are no nearby clusters in the regions covered.  The
rich cluster A\,0781 at z$=0.298$ is however located within field
2, and so the two affected fields closest to the centre of the
cluster were not included among the thirteen.

We used data from the thirteen control fields to estimate the
background counts and the fluctuation amplitude as in Bernstein et
al. (\citeyear{ber95}).  In this case the background counts were
estimated as the mean of the control field counts (Eq. 1 Bernstein et
al. \citeyear{ber95}), and the fluctuations as the rms of the counts
in each control field respect to the mean estimated in the all area
(Eq. 2 Bernstein et al. \citeyear{ber95}). 

\begin{figure*} 
\centerline{{\resizebox{\hsize}{!}{\includegraphics[angle=-90]{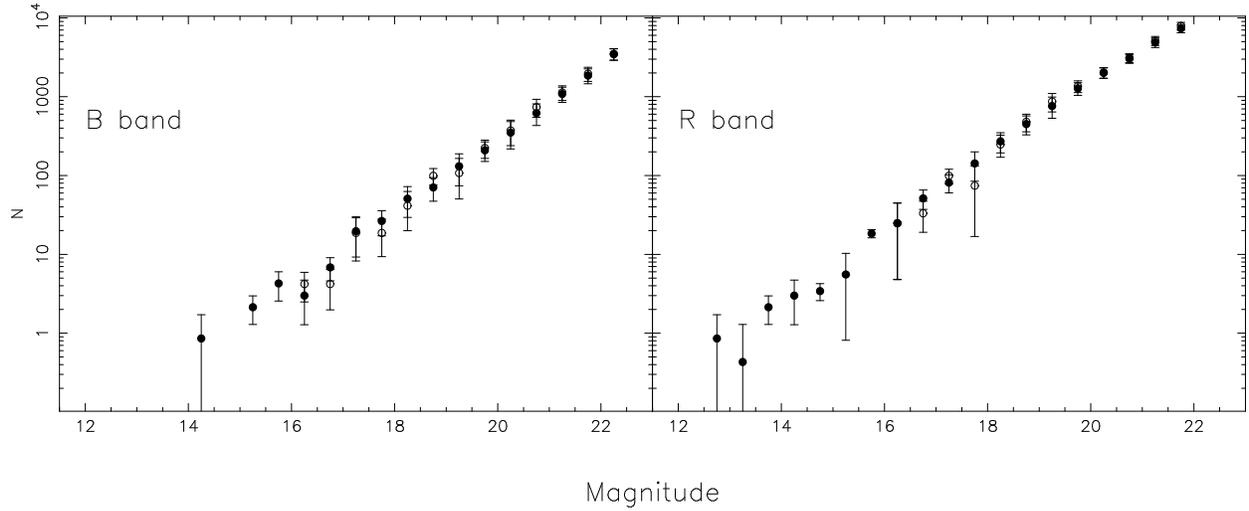}}}}
\caption{Comparison of galaxy counts obtained from the DLS (filled
circles) and ESO-Sculptur Survey (open circles). The counts are
normalized to the total area covered by the SOS data and are per half
magnitude. }
\label{confronto}
\end{figure*}

The galaxy number-magnitude counts obtained from the DLS data were
found to be consistent with those from the literature (e.g., Arnouts et
al. \citeyear{arn97}) for the same passbands (see
Fig.~\ref{confronto}).  An estimate for the total effect from the
cosmic variance from the fluctuations of galaxy counts among the
thirteen fields confirmed that, when combined, the obtained number
magnitude counts are robust against cosmic variance.

The counts of SSC galaxies were defined as the difference between the
counts detected in the supercluster fields and those estimated for the
background (Eq. 3 Bernstein et al. \citeyear{ber95}). By considering
this definition, the uncertainties were measured as the sum in
quadrature of fluctuation in the background and in the supercluster
counts (Eq. 4 Bernstein et al. \citeyear{ber95}).

\begin{figure*} 
\centerline{{\resizebox{7cm}{!}{\includegraphics[angle=-90]{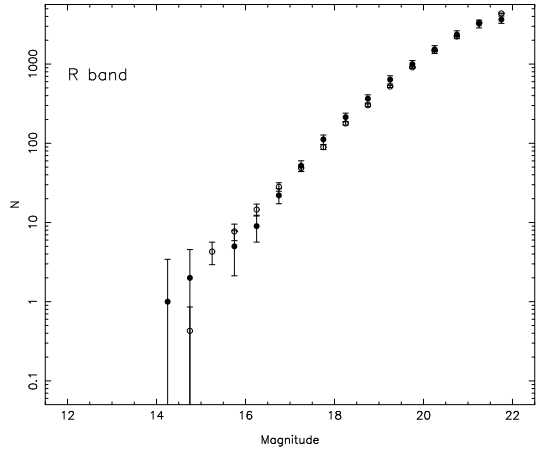}}}}
\caption{Comparison of R-band counts for galaxies redder than the red
sequence (see text) obtained from the SSC (filled circles) and the
DLS (open circles). The counts are normalized to the total area
covered by the SOS data and are per half magnitude. }
\label{confrontored}
\end{figure*}

In order to avoid the background counts taken from the DLS being
too low, we make an additional comparison between SOS and DLS data. We
selected galaxies 3 $\sigma$ redder than the observed red sequence
(Eq.~\ref{CM} and Eq.~\ref{sigmaCM}) in the SOS.  Since galaxies
redward of the sequence should be almost all background galaxies we
compared these counts with those obtained for the DLS control fields
applying the same colour cut. Figure~\ref{confrontored} shows that
these counts are consistent.

\subsection{The total luminosity functions}
\label{sec:62}

Figure \ref{totLF} shows the LFs in B and R bands for galaxies over
the whole 2 deg$^2$ SOS area, covering the SSC core (Fig. 1). The
parameters of the fit are reported in Table~\ref{fitsLF}.  The
weighted parametric fit of a S function (dashed lines in
Fig.~\ref{totLF}) is unable to describe the observed changes in slope
of the LF at faint magnitudes, in particular the dips apparent at B
$\sim$ 17.5 (M$_\mathrm{B} \sim$ - 19.3) and R $\sim$ 17.0
(M$_\mathrm{R} \sim$ - 19.8) and the clear upturn in the counts for
galaxies fainter than B and R $\sim$ 18 mag, apparent in
Fig.~\ref{totLF}. To successfully model these changes in slope
requires a composite G+S LF (continuous line, Fig.~\ref{totLF}), which
represent the data distribution significantly better in both B
(P($\chi^2>\chi^2_\nu$)=97\% against P($\chi^2>\chi^2_\nu$)=0.08\%)
and R bands (P($\chi^2>\chi^2_\nu$)=95\% against
P($\chi^2>\chi^2_\nu$)=23\%). The S function fails most dramatically
to describe the upturn in the galaxy counts at faint magnitudes, as
demonstrated by the composite R-band faint-end slope being -1.62 as
opposed to the S slope of -1.26.

An upper limit to the background counts could be set by using the
counts for galaxies in the SOS region with density less than 0.5,
covering an area of $\sim$~0.5 deg$^2$. The obtained LFs are
consistent with those obtained by using DLS counts, but the error bars
are too large to make any definitive conclusion on the faint-end part
of the luminosity function.

\subsection{The effect of environment}
\label{sec:63}

\begin{figure*} 
\centerline{{\resizebox{\hsize}{!}{\includegraphics[angle=-90]{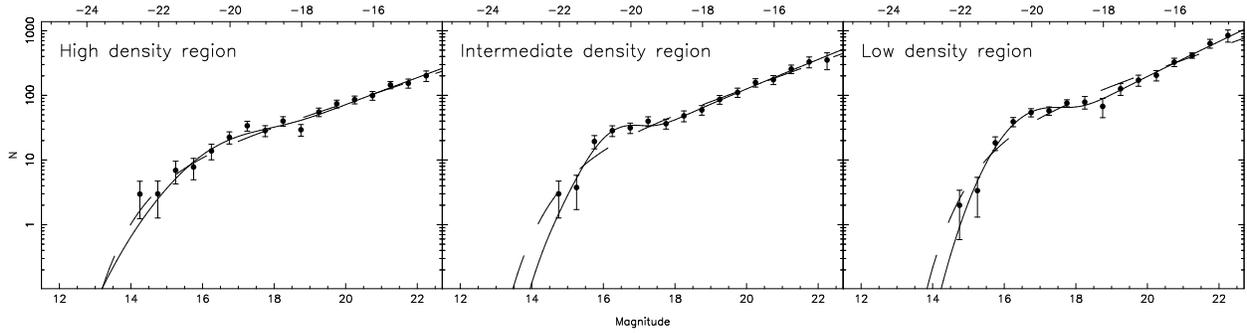}}}}

\caption{The B-band LFs of galaxies in the three cluster regions
corresponding to high-, intermediate- and low-density
environments. Dashed and continuous lines represent the fit with a
S and a G+S respectively. The counts are per half magnitudes.}
\label{Blum}
\end{figure*}

\begin{figure*} 
\centerline{{\resizebox{\hsize}{!}{\includegraphics[angle=-90]{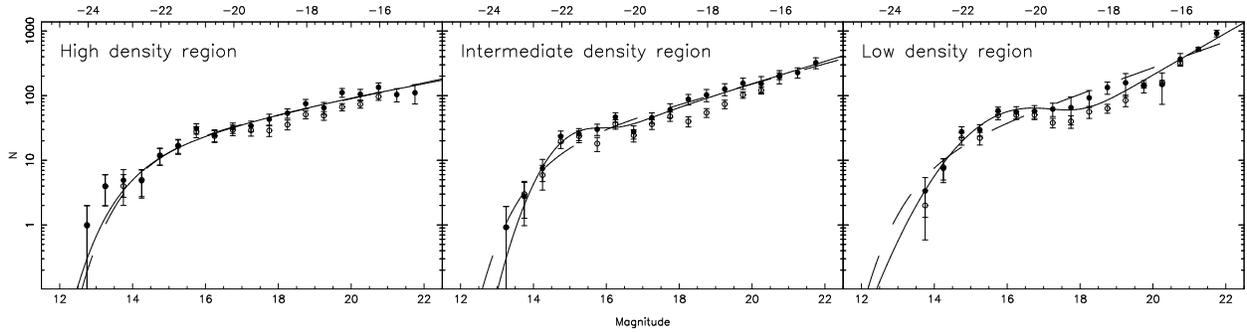}}}}
\caption{The R-band LFs of galaxies in the three regions
corresponding to high-, intermediate- and low-density
environments. Filled circles represent counts obtained from the
photometric catalogue with a statistical background subtraction, open
circles are the counts of galaxies with R$<$21 belonging to the red
sequence of the CM relation (see Sect.~\ref{sec:63}). Dashed and
continuous lines represent the fit with a S and a
G+S function respectively. The counts are per half magnitudes.}
\label{Rlum}
\end{figure*}

Figures \ref{Blum} and \ref{Rlum} show the B- and R-band galaxy LFs in
the high-, intermediate- and low-density regions, covering areas
of $\sim$~0.118, $\sim$~0.344, $\sim$~1.125 deg$^2$,
respectively. Each LF was modelled by a weighted parametric fit to S
(dashed lines) and to composite G+S functions (continuous lines). The
best-fit values are listed in Table~\ref{fitsLF}.

According to the $\chi^2$ statistics in both bands the fit with a
S function can be rejected in the low-density region, the LFs
showing a bimodal behaviour due to the presence of a dip and an upturn
for faint galaxies, that cannot be fitted by using a single
function. In the intermediate-density region, although the S function
cannot be rejected, its fit gives a worse representation of the global
distribution of data compared with a composite function (
P($\chi^2>\chi^2_\nu$) $\sim$ 20\% against P($\chi^2>\chi^2_\nu$)
$\sim$ 70-80\%). On the other hand, in the high-density region the fit
with a S function is more suitable.

\begin{figure}
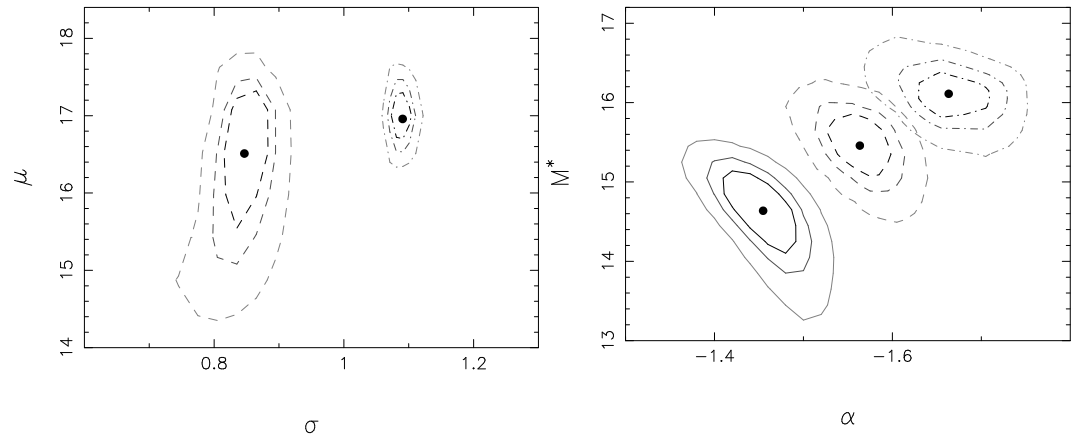

\hbox{
{\resizebox{7cm}{!}{\includegraphics[angle=-90]{figure7a.ps}}}
\hfill
{\resizebox{7cm}{!}{\includegraphics[angle=-90]{figure7b.ps}}}
}
\caption{The 1, 2 and $3\sigma$ confidence levels for the B-band
best-fitting Gaussian (left panel) and Schechter (right panel)
parameters for the three cluster regions corresponding to high- (solid
contours), intermediate- (dashed) and low-density (dot-dashed)
environments. Contours in the high-density region are obtained by
fitting data with a S function.}
\label{contB}
\end{figure}

\begin{figure}
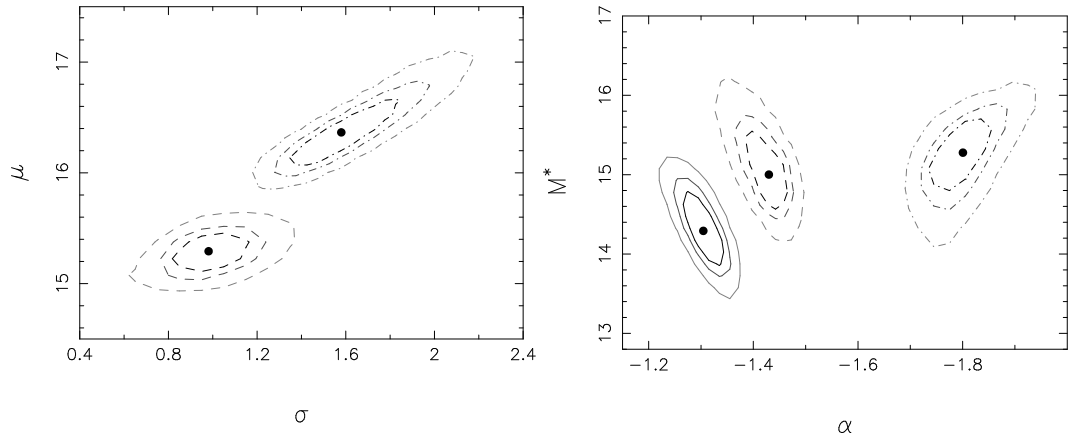

\hbox{
{\resizebox{7cm}{!}{\includegraphics[angle=-90]{figure8a.ps}}}
\hfill
{\resizebox{7cm}{!}{\includegraphics[angle=-90]{figure8b.ps}}}
}
\caption{The 1, 2 and $3\sigma$ confidence levels for the R-band
best-fitting Gaussian (left panel) and Schechter parameters (right
panel) for the three cluster regions corresponding to high- (solid
contours), intermediate- (dashed) and low-density (dot-dashed)
environments. Contours in the high-density region are obtained by
fitting data with a S function.}
\label{contR}
\end{figure}

\begin{figure}
\centerline{{\resizebox{\hsize}{!}{\includegraphics[angle=-90]{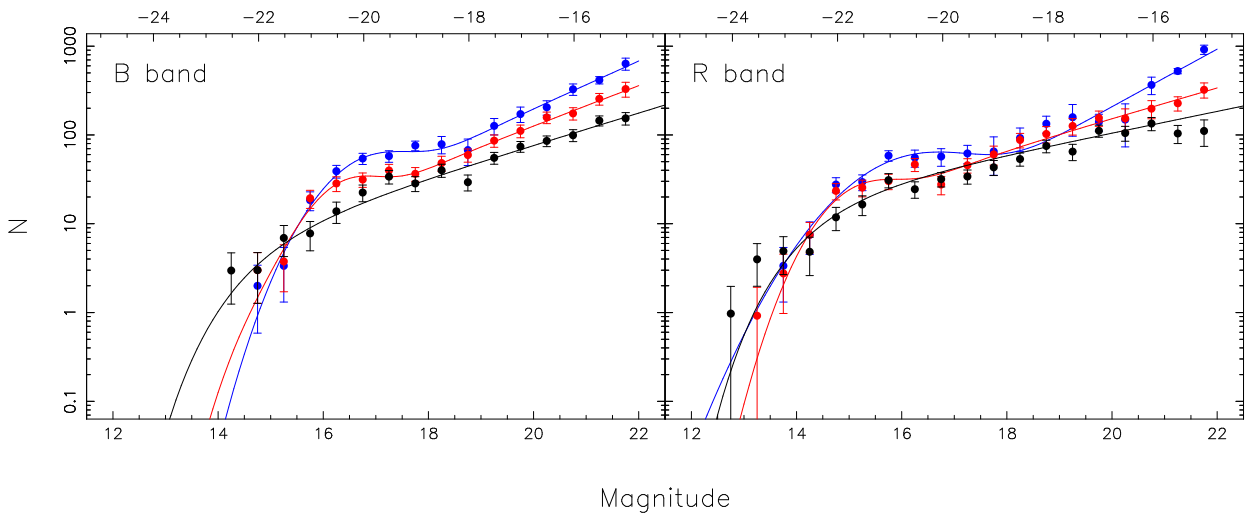}}}}
\caption{The B- (left panel) and R-band (right panel) LFs of galaxies
in the three cluster regions corresponding to high- (black),
intermediate- (red) and low-density (blue) environments. Continuous
lines represent the best fit. The counts are per half magnitudes.}
\label{LF1plot}
\end{figure}

Figures \ref{contB} and \ref{contR} show the confidence contours of
the best fitting functions for B and and R band, respectively, for the
three density regions. The faint-end slope becomes significantly
steeper from high- to low-density environments varying from -1.46 to
-1.66 in B and from -1.30 to -1.80 in R band, being inconsistent at
more than 3$\sigma$ confidence level (c.l.) in both bands (right panel
Figs.~\ref{contB} and \ref{contR}).  Also the bright-end LF is
inconsistent at more than 3$\sigma$ c.l. in both bands, indicating
that also the bright galaxy populations in the SSC depend on the
environment. We note that the shape of the LFs vary dramatically from
high- to low-density regions in both bands (see Fig.~\ref{LF1plot} for
a direct comparison), demonstrating the strong effects of
supercluster environment in low-density regions.

\subsection{Red and blue galaxies}
\label{sec:64}

In order to further investigate the processes responsible for shape
the galaxy LF, we divided the galaxies into red and blue according to
their location with respect to the colour magnitude relation.

We determined the colour-magnitude (CM) relation by performing 100
Monte-Carlo realizations of the supercluster populations and fitting
the photometric data up to R=19 by using the biweight algorithm of
Beers, Flynn, \& Gebhardt (\citeyear{bee90}), obtaining:

\begin{equation}
\mathrm{(B-R)_{CM}= 2.3312 - 0.0563 \times R}  \ . \
\label{CM}
\end{equation}

We also evaluated the B-R colour dispersion around the red sequence as
a function of the magnitude, $\sigma$(R). The dispersion around the
sequence is found to be consistent with the relation:

\begin{equation}
\mathrm{\sigma(R)^2= \sigma_{int}^2 + \sigma^2_{(B-R)}(R)} \ , \ 
\label{sigmaCM}
\end{equation}

\noindent
where the intrinsic dispersion $\mathrm{\sigma_{int}^2}$ is equal to
0.0450 mag over the whole magnitude range covered (Haines et
al. \citeyear{hai05}). 

\begin{figure*} 
\centerline{{\resizebox{10cm}{!}{\includegraphics[angle=-90]{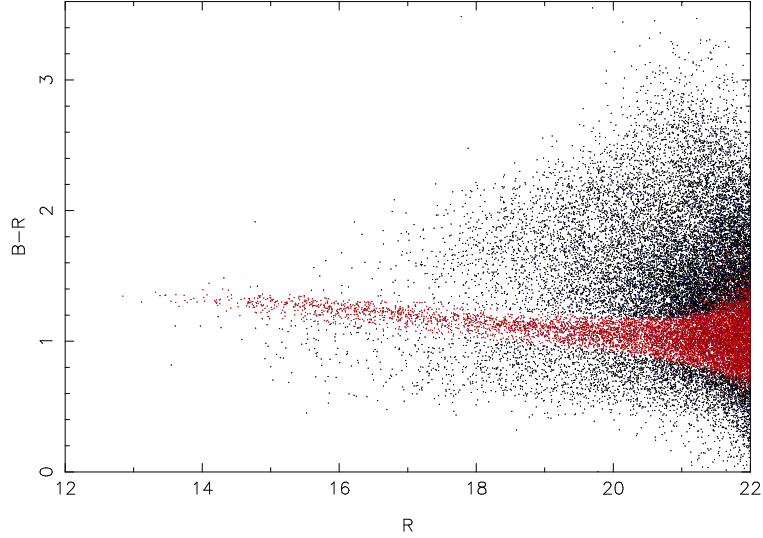}}}}
\caption{B-R vs R CM diagram for all the galaxies up to the
completeness magnitude R=22.0 in the SOS field. Galaxies of the
red sequence (see solid line) are plotted as red points.}
\label{CMfig}
\end{figure*}

In Fig.~\ref{CMfig} the red sequence galaxies are plotted as red
points. We directly compare the counts of galaxies selected on the CM
relation (open circles in Fig.~\ref{Rlum}) with those derived in
Sect.~\ref{sec:63} (filled circles in Fig.~\ref{Rlum}) in order to
exclude projection effects due to background clusters. The counts for
the sequence galaxies were obtained through a statistical background
subtraction, applying the same colour cut of SOS galaxies to those in
the DLS control field. The distributions of red galaxies in the three
different density regions are well described by the total LFs, since
also in this case there is a dip at R$\sim$ 17.0 (M$_\mathrm{R} \sim$
-19.8).

   \begin{table*} 
     \caption[]{Fits to the LFs of blue galaxies. Errors on the
      $\mathrm{M^*}$ and $\alpha$ parameters are shown by the
      confidence contours shown in Fig.~\ref{contbluR}.}

     $$
           \begin{array}{c | c c c | c c }
            \hline
            \noalign{\smallskip}
	    \mathrm{Region} &
            \mathrm{m^*} & \mathrm{M^*} &\alpha & \mathrm{\chi^2_{\nu}} & \mathrm{P(\chi^2>\chi^2_{\nu})}\\
		\noalign{\smallskip}
		\hline
		\hline
	    \noalign{\smallskip}
            \mathrm{high \ \ density} & 16.59 & -20.19 & -1.39 & 0.78& 66.1\% \\
            \mathrm{int \ \ density}  & 14.66 & -22.12 & -1.56 & 0.96& 48.9\% \\
            \mathrm{low \ \ density} & 14.70 & -22.08 & -1.52 & 1.05& 40.0\% \\
            \noalign{\smallskip}
            \hline
            \noalign{\smallskip}
            \hline
         \end{array}
    $$
      \label{fitsLFcol}

 \end{table*}

\begin{figure*} 
\centerline{{\resizebox{\hsize}{!}{\includegraphics[angle=-90]{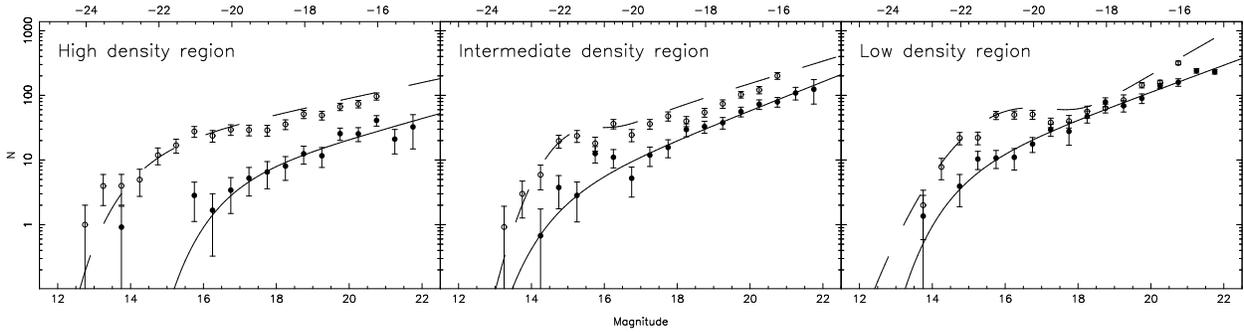}}}}
\caption{The R-band LFs of red (open circles) and blue galaxies
(filled circles) in the three cluster regions corresponding to high-,
intermediate- and low-density environments. Continuous and dashed
lines represent the best fit for blue and supercluster galaxies
respectively. The counts are per half magnitudes.}
\label{Bblulum}
\end{figure*}

\begin{figure}
\centerline{{\resizebox{7cm}{!}{\includegraphics[angle=-90]{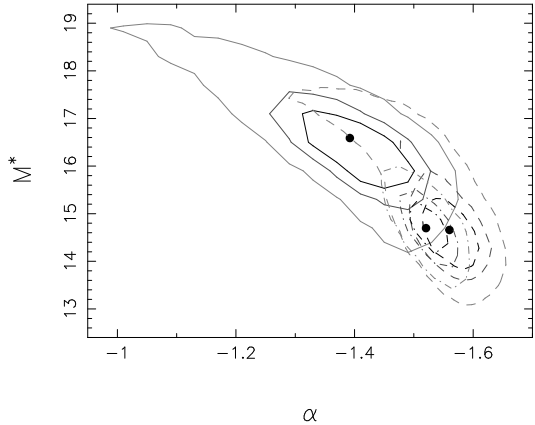}}}}
\caption{The 1, 2 and $3\sigma$ confidence levels for the R-band
best-fitting Schechter parameters for the blue galaxies in the three cluster regions
corresponding to high- (solid contours), intermediate- (dashed) and
low-density (dot-dashed) environments.}
\label{contbluR}
\end{figure}

We also selected the blue supercluster galaxies considering the
galaxies 3$\sigma$ bluer than the CM relation. Figure~\ref{Bblulum}
shows the LFs obtained for the red sequence and blue galaxy population
in high-, intermediate- and low-density regions. The blue galaxy
LFs were obtained through a statistical background subtraction,
applying the same colour cut of the supercluster galaxies to those in
the DLS control fields. In contrast to the red sequence galaxies, the
blue galaxy LFs are well described by a S function and
do not vary with the density (see contours in
Fig.~\ref{contbluR}). This indicates that the blue galaxies represent a
population that have not yet interacted with the supercluster environment.

\section{Summary and discussion}
\label{sec:7}

We have presented a detailed analysis of the LFs for galaxies in the
SSC core. All the luminosity functions were calculated through a
weighted parametric fit of a single Schechter function and a composite
function, given by the sum of a Gaussian for the bright-end and a
Schechter for the faint-end of the LF. The main results of our
analysis are the following.

\begin{description}

\item[-] The LFs in the whole SOS area have a bimodal behaviour both
in B and R band. The weighted parametric fit of a S function is unable
to describe the observed LF at faint magnitudes, in particular the
dips apparent at B $\sim$ 17.5 (M$_\mathrm{B} \sim$ - 19.3) and R
$\sim$ 17.0 (M$_\mathrm{R} \sim$ - 19.8) and the clear upturn for
galaxies fainter than 18 mag. To successfully model these dip and
changes in slope a composite G+S LF is required.

\item[-] By deriving the LFs in regions with different local surface
densities of R$<21.0$ galaxies we showed that, as observed in the LFs
of the whole field, a dip is present at M$_{\mathrm{R}} \sim$ -19.8
for LFs in intermediate- and low-density regions, while for the
high-density region, the data are well represented by the S
function. Moreover the faint-end slope, $\alpha$, shows a strong
dependence on environment, becoming steeper at $>3\sigma$ significance
level from high- ($\alpha_{\mathrm{B}}$ = -1.46, $\alpha_{\mathrm{R}}$
= -1.30) to low-density environments ($\alpha_{\mathrm{B}}$ = -1.66,
$\alpha_{\mathrm{R}}$ = -1.80) in both bands.

\item[-] We derived the LFs separately for red and blue galaxy
populations according to their B-R colours. The LFs of these two
populations show a very different behaviour. In fact differently from
the red sequence galaxy counts that are very similar to those obtained
with a statistical background subtraction, the blue galaxy LFs are
well described by a S function and do not vary with the density.

\end{description}

These results confirm and extend those of MGP94 who found a peak in
the number of galaxies at ${\it b}=18$ and suggested that the Abell
function is a better representation of the integral counts than the
S function. However, their optical LF is limited at galaxies
three magnitudes brighter than those analysed in the present work,
preventing the determination of the steepening of the LF faint-end
and a more clear definition of the LF shape.  On the other hand, MGP94
also noted that the CM red sequence galaxies show the broad peak
at bright magnitudes in agreement with our findings. 
 
The bimodality of the galaxy LF is commonly observed for rich clusters
(e.g., Yagi et al. \citeyear{yag02}; Mercurio et al. \citeyear{mer03}),
and using data from the RASS-SDSS galaxy cluster survey, Popesso et
al. (\citeyear{pop05}) find a similar variation of the LF with
environment to that observed here, but using cluster-centric radius
rather than local density (e.g., Haines et al. \citeyear{hai04}) as a
proxy for environment. This observed bimodality and its variation with
environment can be best accomodated in a scenario where bright and
faint galaxy populations have followed different evolution histories.

The SDSS and 2dFGRS surveys have indicated that the evolution of
bright galaxies is strongly dependent on environment as measured by
their local density, yet is independent of the richness of the
structure to which the galaxy is bound, indicating that mechanisms
such as merging or suffocation play a dominant role in transforming
galaxies, rather than harassment or ram pressure stripping (G\'omez et
al. \citeyear{gom03}; Tanaka et al.  \citeyear{tan04}). However, it is
difficult to reconcile the dramatic deviations from the S function
observed for intermediate- and low-density regions with the
transformation of field galaxies being due to just merging or
suffocation, neither of which should alter the shape of LF, whilst
they can be explained more easily by a scenario involving mass loss of
low-luminosity galaxies.

One such mechanism is galaxy harassment (Moore et
al. \citeyear{moo96}, \citeyear{moo98}), whereby repeated close
($<$50\,kpc) high-velocity ($>$1000\,km\,s$^{-1}$) encounters with
bright galaxies and the cluster's tidal field cause impulsive
gravitational shocks that damage the fragile disks of late-type
spirals. The cumulative effect of these shocks is the transformation
of late-type spirals to spheroidal galaxies over a period of several
Gyr.  An important aspect of galaxy harassment is that it has
virtually no effect on systems as dense as giant elliptical or spiral
bulges, and hence only pure disk systems (Sc or later) are
affected. While these galaxies make up the vast majority of the faint
(M$>$M$^*$+2) cluster galaxy population at $\mathrm{z}\gtrsim0.4$,
they become rarer exponentially at brighter magnitudes. The
encounters can drive the bulk of the dark matter and 20--75\% of the
stars over the tidal radius of the harassed galaxy, whereas in
contrast the bulk of the gas collapses inward, and is consumed in a
nuclear starburst. The combined results of these effects is a dimming
of the harassed galaxy by $\sim2$ magnitudes due to mass loss and
passive aging of the remaining stars. These remnants are apparent in
present day clusters as dwarf spheroids which often show blue cores
suggesting nuclear star-formation, as well as remnant disk and bar
components (Graham, Jerjen \& Guzman \citeyear{gra03}), and signs of
rotational support (de Rijcke et al. \citeyear{der01}).

In agreement with the recent work by Popesso et al.
(\citeyear{pop05}) we suggest that the observed dip at
$M_{R}\sim-19.8$ as well as the strong dependence on environment shown
by the faint-end slope in the cluster galaxy luminosity can be
explained naturally as the consequence of galaxy harassment.

Alternative mechanisms such as ram-pressure stripping by the ICM or
tidal stripping can effect the galaxy population only in the cluster
cores, which appears inconsistent with our observation that the dip is
greatest in the low-density regions 1--2\,Mpc from the nearest
cluster. However, given the high infall velocities, any galaxy
encountering the ICM is likely to be stripped rapidly of their gas,
bringing star-formation to a swift halt. Given the high infall
velocities, and the typical highly eccentric orbits of cluster
galaxies, the low- and intermediate-density regions are likely to
contain a significant fraction of galaxies that have already
encountered the dense ICM.  In high-density regions, high-velocity
dispersions inhibit merging processes (e.g., Mihos \citeyear{mih04}),
hence it is unlikely that dwarf galaxies merge to produce bigger
galaxies at the cluster centres. The most likely explanation for the
lack of dwarf galaxies near the centre is tidal or collisional
disruption of the dwarf galaxies.

This interpretation is also confirmed when analysing separately
red sequence galaxies. In fact the red galaxy counts exhibit a
behaviour similar to those of the LFs obtained with a statistical
background subtraction, confirming the excess of dwarf early type
galaxies. Moreover, differently from red sequence galaxies, the blue
galaxy LFs are well described by a S function with a slope $\alpha
\sim$ -1.50 and do not vary with density. This slope is consistent
with those recently derived by Blanton et al. (\citeyear{bla05}) and
Madgwick et al. (\citeyear{mad02}) for field SDSS and 2dF galaxies
respectively. This suggest that the observed blue galaxy population is
characterized by infalling galaxies that have not yet interacted with
the super cluster environment and transformed by the harassment mechanism.

In a forthcoming paper (Haines et al. \citeyear{hai05}) we will
investigate in detail the distribution of red and blue galaxies in the
SSC environment.

\section*{Acknowledgments}   

We thank the Deep Lens Survey and NOAO, who provided the galaxy counts
used to derive the LFs. We thank the anonymous referee for useful
comments.  AM is supported by the Regione Campania (L.R. 05/02)
project {\it `Evolution of Normal and Active Galaxies'} and by the
Italian Ministry of Education, University, and Research (MIUR) grant
COFIN2004020323: {\it The Evolution of Stellar Systems: a Fundamental
Step towards the Scientific Exploitation of VST}.  CPH , AG and NK
acknowledge the financial supports provided through the European
Community's Human Potential Program, under contract HPRN-CT-2002-0031
SISCO. NK and GC are partially supported by the Italian Ministry of
Education, University, and Research (MIUR) grant COFIN2003020150: {\it
Evolution of Galaxies and Cosmic Structures after the Dark Age:
Observational Study} and grant COFIN200420323: {\it The Evolution of
Stellar Systems: a Fundamental Step towards the Scientific Explotation
of VST}, respectevely.

\label{lastpage}
\end{document}